\documentclass[prx, tightenlines,10pt,notitlepage,nofootinbib,superscriptaddress,twocolumn]{revtex4-1}
\usepackage[utf8]{inputenc}
\usepackage{natbib}
\usepackage{graphicx}
\usepackage{epstopdf}
\usepackage{amsmath}
\usepackage{lineno}
\usepackage{comment}
\usepackage{graphicx, nicefrac}
\usepackage{float}
\usepackage{subfigure}
\usepackage{hyperref}
\usepackage[english]{babel}

\hypersetup{
    pdfpagemode=FullScreen,
    }

\begin{document}
\title{LightSolver - A New Quantum-inspired Solver Cracks the 3-Regular 3-XORSAT Challenge}


\author{Idan Meirzada}
\email[Corresponding Author:\,]{idan@lightsolver.com}
\affiliation{LightSolver LTD., Tel Aviv, Israel}
\author{Assaf Kalinski}
\affiliation{LightSolver LTD., Tel Aviv, Israel}
\author{Dov Furman}
\affiliation{LightSolver LTD., Tel Aviv, Israel}
\author{Tsafrir Armon}
\affiliation{LightSolver LTD., Tel Aviv, Israel}
\author{Talya Vaknin}
\affiliation{LightSolver LTD., Tel Aviv, Israel}
\author{Harel Primack}
\affiliation{LightSolver LTD., Tel Aviv, Israel}
\author{Chene Tradonsky}
\affiliation{LightSolver LTD., Tel Aviv, Israel}
\author{Ruti Ben-Shlomi}
\email[Corresponding Author:\,]{ruti@lightsolver.com}
\affiliation{LightSolver LTD., Tel Aviv, Israel}

\begin{abstract}

The increasing complexity of required computational tasks alongside the inherent limitations in conventional computing calls for disruptive innovation. LightSolver devised a new quantum-inspired computing paradigm, which utilizes an all-optical platform for solving hard optimization problems. 
In this work, LightSolver introduces its digital simulator and joins the 3-Regular 3-XORSAT (3R3X) challenge, which aims to map the best available state-of-the-art classical and quantum solvers. So far, the challenge has resulted in a clear exponential barrier in terms of time-to-solution (TTS), preventing the inspected platforms from solving problems larger than a few hundred variables. LightSolver's simulator is the first to break the exponential barrier, outperforming both classical and quantum platforms by several orders-of-magnitude and extending the maximal problem size to more than 16,000 variables.

\end{abstract}

\maketitle

\section*{Introduction}\,

Solving optimization problems holds the potential for numerous advancements in a myriad of applications in both research and industry, from autonomous transportation to resource management \cite{Integer_and_Combinatorial_Optimization}. 
In recent years, this challenge inspired an increasing number of research groups and companies to develop dedicated techniques, by designing innovative heuristics \cite{tabu_glover,Simulated_Quantum_Annealing,SBM_Combinatorial_optimization_by_simulating_adiabatic_bifurcations}, utilizing sophisticated architectures \cite{NN} or realizing novel, cutting-edge hardware \cite{fujitsu_tsukamoto2017accelerator,YamamotoNat,YamSci,MEM_NP_complete_problems}. These solutions offer simplicity and availability, flexible connectivity or short convergence times. However, they often depend on cooling to low temperatures, suffer from limited connectivity or dynamic range, or require vast amounts of energy when applied to real-life problems. 
\\
One year ago, two research groups at the University of Southern California launched a computing challenge to compare the actual performances of the best available state-of-the-art solvers  \cite{3-xor-sat}. This work revealed the differences, as well as certain similarities, between their respective performances. These results were presented, together with other problems of simpler configurations, in a recent review article \cite{McMahon_Ising_machines}.

In this work, we introduce LightSolver, a novel platform that implements a new concept for solving hard optimization problems, based on a spatially coupled laser array. 
We provide a first glance into the performance of this platform by testing a digital simulation of the optical platform on similar 3R3X instances, and compare its results to the solving platforms detailed in the challenge \cite{3-xor-sat} while also extending the maximal problem size from 640 variables to 16,384 variables. 

In the next sections, we will introduce our optical platform, the 3R3X problem, and examine the performance of the LightSolver digital simulator on 3R3X instances in the absence and in the presence of noise. We conclude by discussing these results and their implications for the optical platform. 
\\

\section*{LightSolver's Analog Platform}

Similar to many other solving platforms, LightSolver utilizes the flexible nature of the Ising model to formulate optimization problems into a Hamiltonian form, consisting of linear and quadratic terms \cite{Lucas}. This process allows a single solving method to address a wide variety of problems, instead of separately tailoring a different process for each problem category. 
\newline The Ising Hamiltonian takes the following structure: 
\begin{equation}
H = \sum_{i=1}^n h_is_i + \sum_{i, j = 1}^n J_{ij}s_is_j
\end{equation} 
where $s_i \in \{-1,1\}$, $h_i$ defines the external field, $J_{ij}$ defines the interactions and $n$ stands for the number of variables. 
\\ \\ 
In the physical (optical) layer, the spin state is encoded in the lasers' relative phases, thus promoting a representation of a binary Ising Hamiltonian via a continuous problem. The spins interact by diffracting light from each laser to all other lasers in a controllable manner, using a unique optical coupler composed of programmable diffractive elements and additional optical components. The coupler controls the interaction between all laser pairs, with a dynamical range of up to 8 bits. This design allows for a full connectivity between all lasers, facilitating pairwise all-to-all high-resolution spin interactions on a desktop-size device, operating at room temperature, while requiring only a modest amount of energy ($ < 1kW)$. 
\\ \\ 
Unlike other experimental realizations, LightSolver's platform implements an expanded version of the XY model, due to continuity in both phase and amplitude. This, on the one hand, requires constructing a novel formulation and algorithm to tackle binary optimization problems, such as Ising. On the other hand, the continuous nature allows modifying the formulations to accommodate different requirements according to problem type (e.g. SAT, Max Cut, Max Clique, etc.), objective attributes (e.g. size, connectivity, dynamic range, internal structure) or solution preferences (speed, quality, diversity). The algorithm and formulations remain beyond the scope of this work. 

\section*{Method}


The 3R3X problem \cite{Hiding_Solutions,Being_Glassy} serves as a test problem for optimization solvers, allowing the generation of a rough energy topography while knowing in advance the problem's optimal solution. These problems consist of $n/2$ exclusive-OR clauses, each of which includes three binary literals, so that satisfying a clause requires one or three literals to be True. Each literal appears exactly in three clauses, resulting in a symmetric connectivity, forming a 'golf-course' shaped energy landscape. The $n$-variable Ising Hamiltonians used for this work correspond to $n/2$ 3R3X randomly generated problem instances, ranging from $n = 32$ to $n = 16,384$. For more details on the nature of the problem, the reader is referred to Ref.~\cite{PhysRevApplied.12.011003}.
\\\\
Since a variety of state-of-the-art solving platforms and algorithms already addressed this problem, it is a solid candidate for comparison. Existing solvers include application-specific integrated circuit (ASIC) Digital Annealer Unit by Fujitsu (DAU) \cite{fuj1}, superconducting qubits based annealer by D-Wave (Advantage model) \cite{dwave1}, tailor-made SAT solver running on GPU (SAT on GPU) \cite{SATonGPU}, Virtual MemComputing Machine (MEM) \cite{MEM}, and two classical algorithms - Simulated Bifurcation Machine by Toshiba (SBM) \cite{sbm1} and Parallel Tempering (PT) \cite{PT1,PT2}, implemented on single GPU and CPU cores, correspondingly.
\\\\
To add confidence to the veracity of the results, the performance of the Lightsolver simulator was tested and analyzed by an external party (namely, one of the co-authors of Ref.~\cite{3-xor-sat}), asking them to perform the simulation on our simulator's alpha version. The following sections portray the obtained results. 
\\\\



\section*{Results}

The LightSolver simulator ran on a single g4dn.4xlarge (64GB RAM) machine through an Amazon cloud service. To reliably predict the asymptotic behavior of the algorithm, only one initial state evolved in each run, although the simulator supports running many initial states in parallel using the GPU. 
Each of the 23 problem sizes examined in this work comprises 25 randomly generated instances. For each instance, the simulation ran until it reached the ground-state or the maximal number of steps defined (100,000 for noise-free simulations and 500,000 for noise-injected simulations), and the TTS was calculated. 
\\\\
The TTS of a single initial state depends on the simulation time ($t_f$) and the probability to reach the ground state ($p_i$), and is calculated in the following way \cite{TTS}:

\begin{centering}
    \begin{equation}
        TTS = \underset{t_f}{min}\biggl \langle t_f \frac{ln(1-0.99)}{ln[1-p_i(t_f)]} \biggr \rangle.
    \end{equation}
\end{centering}

\subsection{Noise-free simulations}

\begin{figure*}
    {\includegraphics[width=0.75\linewidth]{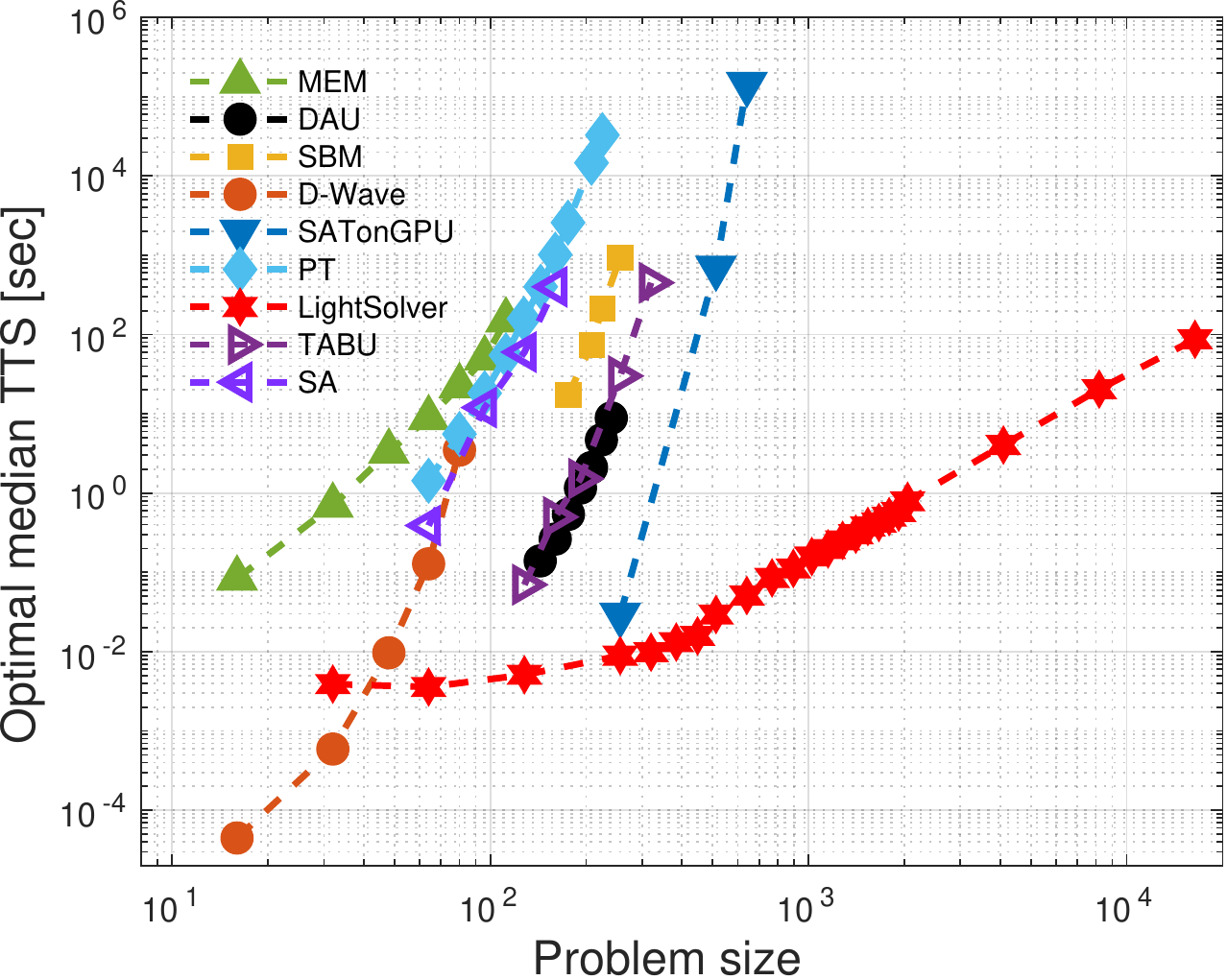}}
    \caption{Optimal TTS of different solving platforms for 3-XOR-SAT problem as measured in \cite{3-xor-sat}, including LightSolver simulator (red stars), as well as tabu search and simulated annealing (dark and light purple triangles, respectively), for up to $n = 16,384$ variables. The LightSolver simulator displays a power-law scaling of the form $TTS \sim n^k$, where the power $k = 2.31 \pm 0.03$, is solving a 16,384 variable problem in less than 100 seconds. Tabu search shows similar behavior to the Fujitsu DAU solver, while simulated annealing performs slightly better than PT.}
\label{fig:regular}
\end{figure*}

Figure~\ref{fig:regular} depicts the LightSolver digital simulator's performance, and is the main result of this paper. The figure presents the TTS for the 3R3X problem as a function of problem size, from $n = 32$ to $n = 16,384$ variables, on a log-log scale. The red stars correspond to LightSolver's simulator, comparing its performance to other state of the art platforms benchmarked in \cite{3-xor-sat}, as well as known heuristics, namely Tabu Search and Simulated Annealing. 
\\
\\

LightSolver's simulator demonstrates a power-law scaling of the form $TTS \sim n^k$, where the power $k = 2.31 \pm 0.03$. For reference, all of the other solving platforms demonstrated exponential scaling of the form $TTS \sim 10^{\alpha n}$, with the pre-factor $\alpha$ varying from 0.0171 (SAT on GPU) to 0.08 (D-Wave advantage), resulting in time-to-solution $TTS > 10^4$ seconds for a 640 variable problem for the best performing solver. Thus, despite falling short when facing smaller instances, LightSolver's simulator proves superior and exhibits 1-5 orders of magnitude speed-up for every tested problem larger than $ n > 64 $ variables, while evolving only a single initial state.

\subsection{Noise-injected simulations}
Due to its analog nature, the optical platform may suffer from substantial noise sources, such as thermal fluctuations, and optical and electrical instabilities. These would interfere with the 'natural' evolution of the laser phases and amplitudes of the coupled lasers, impeding the system's ability to effectively solve the problem. 
Thus, to better predict the optical platform's performance, we introduced noise to the simulation in the following way: at each step (which corresponds to a round trip in the optical platform) the simulation samples a vector out of a normal distribution around the chosen noise amplitude, and adds it to the current vector. The noise distorts the phases and amplitudes of each laser and thus changes the effective interactions between the Ising spins.

\begin{figure}[tbh]
    {\includegraphics[width=1.0\linewidth]{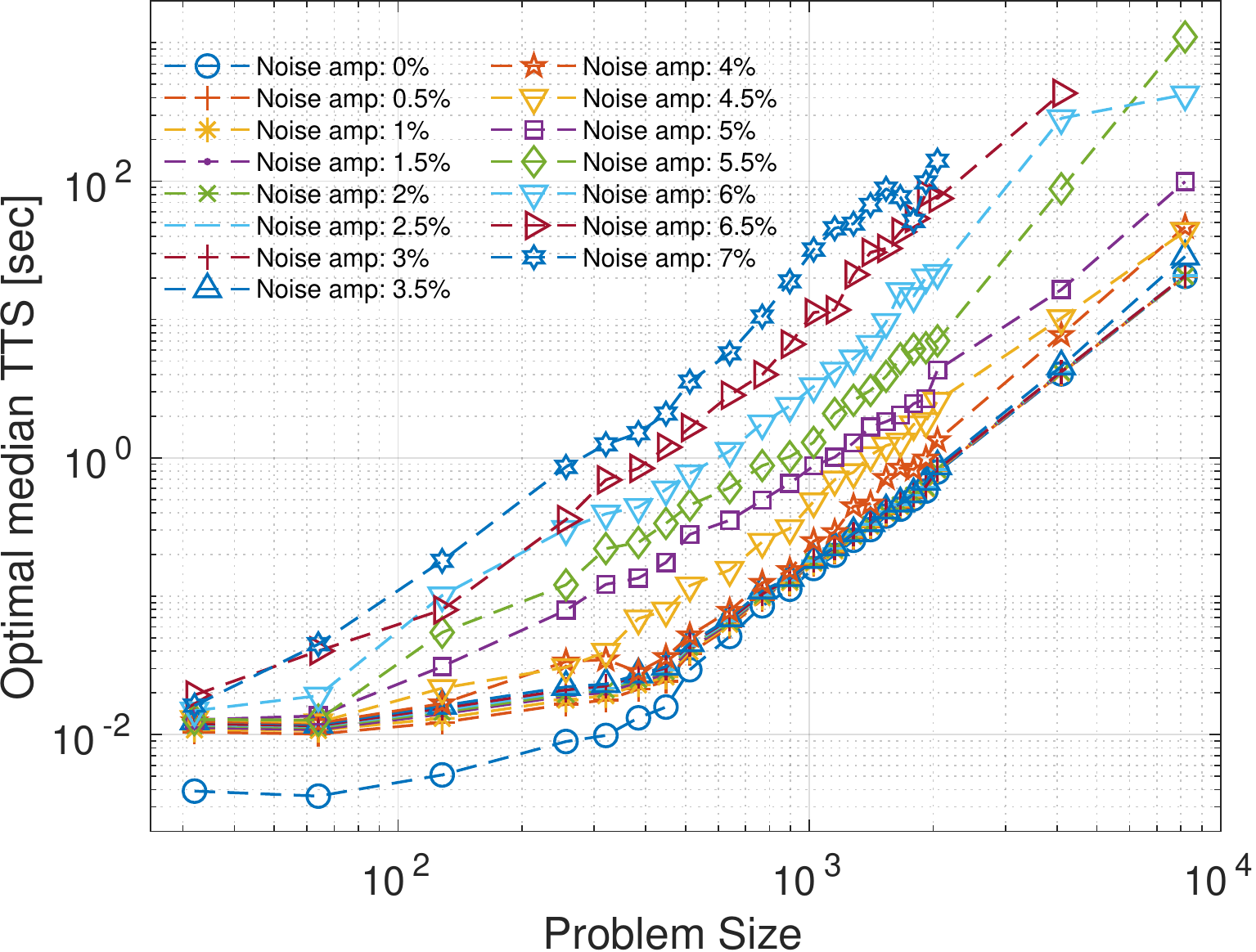}}
    \caption{TTS as a function of problem size, for noise amplitudes ranging from 0 to 7\%. The system seems indifferent to noise levels below 3\%, and maintains similar scaling for all noise amplitudes tested.}
\label{fig:noise}
\end{figure}

Fig. \ref{fig:noise} outlines the LightSolver simulator's TTS as a function of problem size in the presence of noise, for noise amplitudes up to 7\% of the saturated amplitude (an intrinsic property in the simulation limiting the lasers' amplitude according to gain dynamics), for evolutions of up to 500,000 steps. 

The simulation is robust in the presence of noise, demonstrating an almost complete resilience for noise levels of up to 3\% for all problem sizes, and suffering order-of-magnitude increases in TTS only when confronting large noise levels of above 5.5\%. Moreover, the TTS scaling remains constant regardless of the noise levels inspected here, taking the form of a similar power law with the power $k = 2.3 \pm 0.2$. The 'jump' in TTS between noiseless evolution and the rest of the curves at the beginning of the figure stems from noise generation overhead in the simulation code. 

\section*{Discussion \& Conclusions}
The 3R3X problem challenges many state-of-the-art heuristics and physical Ising solvers. Although Gaussian elimination solves this problem with a cubic scaling, so far every Ising solver benchmarked against this problem presented an exponential scaling, with coefficients ranging between $0.0171$ to $0.08$ \cite{3-xor-sat}, resulting in $TTS > 10^4 $ seconds for problems larger than a few hundred variables. 
\\ \\ 
The LightSolver simulator testing indicates a speed-up of several orders of magnitude compared to the benchmarked solvers, revealing a power law scaling, thus allowing solving problems of more than 16,000 spins in less than 100 seconds. Furthermore, we did not exploit all of the parallel capabilities of the simulator (e.g., running many initial states at the same time) to evaluate the asymptotic scaling of the TTS as a function of problem size. Relaxing this restriction would result in better TTS and scaling, especially for evolutions in the presence of noise. 
\\ \\ 
The increase in computation time originates mostly in the emulation of the diffractive elements of the physical hardware on the computing platform (scaling as $n^2$). This implies that although the problems increase in size, their hardness increases very slowly in the eyes of the algorithm. Additionally, the algorithm demonstrates its ability to reach the global minima in the presence of noise, for amplitudes up to 7\%, while keeping a similar scaling.
\\ \\ 
The hardware implementation of the solver consists of a coupled laser array, with an all-to-all connectivity. Thus, the time required to solve a problem on the optical platform depends solely on the number of round trips required, independent of the problem size. The fact that for each noise level the required number of round trips remains almost constant, suggests that the optical solver would scale even better than the simulation, resulting in a speed-up of more than three orders of magnitude compared to the simulation results (provided that the optical solver includes enough spins to contain the entire problem). 
\\ \\ 
Lastly, an important aspect to consider when discussing performance of any solver, concerns the energy consumption required by the proposed solving platform. The suggested analog hardware, simulated in this work, would perform the 'heavy' matrix-vector multiplication using an optical setup, regardless of problem size or connectivity (up to $\sim$ 1000 variables for the first generation), while working at room temperature. Thus, the optical platform will potentially consume a few hundred Watts, comparable to an average desktop.
\\
\section*{Acknowledgements}
We wish to thank Dr. Itay Hen (University of Southern California), for performing the simulations on our alpha version of the simulator and the in-depth analysis of the results. 

\bibliography{ref}

\begin{thebibliography}{23}%
\makeatletter
\providecommand \@ifxundefined [1]{%
 \@ifx{#1\undefined}
}%
\providecommand \@ifnum [1]{%
 \ifnum #1\expandafter \@firstoftwo
 \else \expandafter \@secondoftwo
 \fi
}%
\providecommand \@ifx [1]{%
 \ifx #1\expandafter \@firstoftwo
 \else \expandafter \@secondoftwo
 \fi
}%
\providecommand \natexlab [1]{#1}%
\providecommand \enquote  [1]{``#1''}%
\providecommand \bibnamefont  [1]{#1}%
\providecommand \bibfnamefont [1]{#1}%
\providecommand \citenamefont [1]{#1}%
\providecommand \href@noop [0]{\@secondoftwo}%
\providecommand \href [0]{\begingroup \@sanitize@url \@href}%
\providecommand \@href[1]{\@@startlink{#1}\@@href}%
\providecommand \@@href[1]{\endgroup#1\@@endlink}%
\providecommand \@sanitize@url [0]{\catcode `\\12\catcode `\$12\catcode
  `\&12\catcode `\#12\catcode `\^12\catcode `\_12\catcode `\%12\relax}%
\providecommand \@@startlink[1]{}%
\providecommand \@@endlink[0]{}%
\providecommand \url  [0]{\begingroup\@sanitize@url \@url }%
\providecommand \@url [1]{\endgroup\@href {#1}{\urlprefix }}%
\providecommand \urlprefix  [0]{URL }%
\providecommand \Eprint [0]{\href }%
\providecommand \doibase [0]{http://dx.doi.org/}%
\providecommand \selectlanguage [0]{\@gobble}%
\providecommand \bibinfo  [0]{\@secondoftwo}%
\providecommand \bibfield  [0]{\@secondoftwo}%
\providecommand \translation [1]{[#1]}%
\providecommand \BibitemOpen [0]{}%
\providecommand \bibitemStop [0]{}%
\providecommand \bibitemNoStop [0]{.\EOS\space}%
\providecommand \EOS [0]{\spacefactor3000\relax}%
\providecommand \BibitemShut  [1]{\csname bibitem#1\endcsname}%
\let\auto@bib@innerbib\@empty
\bibitem [{\citenamefont {Hoffman}\ and\ \citenamefont
  {Ralphs}(2013)}]{Integer_and_Combinatorial_Optimization}%
  \BibitemOpen
  \bibfield  {author} {\bibinfo {author} {\bibfnamefont {K.~L.}\ \bibnamefont
  {Hoffman}}\ and\ \bibinfo {author} {\bibfnamefont {T.~K.}\ \bibnamefont
  {Ralphs}},\ }\enquote {\bibinfo {title} {Integer and combinatorial
  optimization},}\ in\ \href {\doibase 10.1007/978-1-4419-1153-7_129} {\emph
  {\bibinfo {booktitle} {Encyclopedia of Operations Research and Management
  Science}}},\ \bibinfo {editor} {edited by\ \bibinfo {editor} {\bibfnamefont
  {S.~I.}\ \bibnamefont {Gass}}\ and\ \bibinfo {editor} {\bibfnamefont {M.~C.}\
  \bibnamefont {Fu}}}\ (\bibinfo  {publisher} {Springer US},\ \bibinfo
  {address} {Boston, MA},\ \bibinfo {year} {2013})\ pp.\ \bibinfo {pages}
  {771--783}\BibitemShut {NoStop}%
\bibitem [{\citenamefont {Glover}(1990)}]{tabu_glover}%
  \BibitemOpen
  \bibfield  {author} {\bibinfo {author} {\bibfnamefont {F.}~\bibnamefont
  {Glover}},\ }\href@noop {} {\bibfield  {journal} {\bibinfo  {journal}
  {Interfaces}\ }\textbf {\bibinfo {volume} {20}},\ \bibinfo {pages} {74}
  (\bibinfo {year} {1990})}\BibitemShut {NoStop}%
\bibitem [{\citenamefont {Finnila}\ \emph {et~al.}(1994)\citenamefont
  {Finnila}, \citenamefont {Gomez}, \citenamefont {Sebenik}, \citenamefont
  {Stenson},\ and\ \citenamefont {Doll}}]{Simulated_Quantum_Annealing}%
  \BibitemOpen
  \bibfield  {author} {\bibinfo {author} {\bibfnamefont {A.}~\bibnamefont
  {Finnila}}, \bibinfo {author} {\bibfnamefont {M.}~\bibnamefont {Gomez}},
  \bibinfo {author} {\bibfnamefont {C.}~\bibnamefont {Sebenik}}, \bibinfo
  {author} {\bibfnamefont {C.}~\bibnamefont {Stenson}}, \ and\ \bibinfo
  {author} {\bibfnamefont {J.}~\bibnamefont {Doll}},\ }\href {\doibase
  https://doi.org/10.1016/0009-2614(94)00117-0} {\bibfield  {journal} {\bibinfo
   {journal} {Chemical Physics Letters}\ }\textbf {\bibinfo {volume} {219}},\
  \bibinfo {pages} {343} (\bibinfo {year} {1994})}\BibitemShut {NoStop}%
\bibitem [{\citenamefont {Goto}\ \emph
  {et~al.}(2019{\natexlab{a}})\citenamefont {Goto}, \citenamefont {Tatsumura},\
  and\ \citenamefont
  {Dixon}}]{SBM_Combinatorial_optimization_by_simulating_adiabatic_bifurcations}%
  \BibitemOpen
  \bibfield  {author} {\bibinfo {author} {\bibfnamefont {H.}~\bibnamefont
  {Goto}}, \bibinfo {author} {\bibfnamefont {K.}~\bibnamefont {Tatsumura}}, \
  and\ \bibinfo {author} {\bibfnamefont {A.~R.}\ \bibnamefont {Dixon}},\ }\href
  {\doibase 10.1126/sciadv.aav2372} {\bibfield  {journal} {\bibinfo  {journal}
  {Science Advances}\ }\textbf {\bibinfo {volume} {5}},\ \bibinfo {pages}
  {eaav2372} (\bibinfo {year} {2019}{\natexlab{a}})}\BibitemShut {NoStop}%
\bibitem [{\citenamefont {Xia}\ \emph {et~al.}(2002)\citenamefont {Xia},
  \citenamefont {Leung},\ and\ \citenamefont {Wang}}]{NN}%
  \BibitemOpen
  \bibfield  {author} {\bibinfo {author} {\bibfnamefont {Y.}~\bibnamefont
  {Xia}}, \bibinfo {author} {\bibfnamefont {H.}~\bibnamefont {Leung}}, \ and\
  \bibinfo {author} {\bibfnamefont {J.}~\bibnamefont {Wang}},\ }\href {\doibase
  10.1109/81.995659} {\bibfield  {journal} {\bibinfo  {journal} {IEEE
  Transactions on Circuits and Systems I: Fundamental Theory and Applications}\
  }\textbf {\bibinfo {volume} {49}},\ \bibinfo {pages} {447} (\bibinfo {year}
  {2002})}\BibitemShut {NoStop}%
\bibitem [{\citenamefont {Tsukamoto}\ \emph
  {et~al.}(2017{\natexlab{a}})\citenamefont {Tsukamoto}, \citenamefont
  {Takatsu}, \citenamefont {Matsubara},\ and\ \citenamefont
  {Tamura}}]{fujitsu_tsukamoto2017accelerator}%
  \BibitemOpen
  \bibfield  {author} {\bibinfo {author} {\bibfnamefont {S.}~\bibnamefont
  {Tsukamoto}}, \bibinfo {author} {\bibfnamefont {M.}~\bibnamefont {Takatsu}},
  \bibinfo {author} {\bibfnamefont {S.}~\bibnamefont {Matsubara}}, \ and\
  \bibinfo {author} {\bibfnamefont {H.}~\bibnamefont {Tamura}},\ }\href@noop {}
  {\bibfield  {journal} {\bibinfo  {journal} {Fujitsu Sci. Tech. J}\ }\textbf
  {\bibinfo {volume} {53}},\ \bibinfo {pages} {8} (\bibinfo {year}
  {2017}{\natexlab{a}})}\BibitemShut {NoStop}%
\bibitem [{\citenamefont {Marandi}\ \emph {et~al.}(2014)\citenamefont
  {Marandi}, \citenamefont {Wang}, \citenamefont {Takata}, \citenamefont
  {Byer},\ and\ \citenamefont {Yamamoto}}]{YamamotoNat}%
  \BibitemOpen
  \bibfield  {author} {\bibinfo {author} {\bibfnamefont {A.}~\bibnamefont
  {Marandi}}, \bibinfo {author} {\bibfnamefont {Z.}~\bibnamefont {Wang}},
  \bibinfo {author} {\bibfnamefont {K.}~\bibnamefont {Takata}}, \bibinfo
  {author} {\bibfnamefont {R.}~\bibnamefont {Byer}}, \ and\ \bibinfo {author}
  {\bibfnamefont {Y.}~\bibnamefont {Yamamoto}},\ }\href {\doibase
  10.1038/nphoton.2014.249} {\bibfield  {journal} {\bibinfo  {journal} {Nature
  Photonics}\ }\textbf {\bibinfo {volume} {8}},\ \bibinfo {pages} {937}
  (\bibinfo {year} {2014})}\BibitemShut {NoStop}%
\bibitem [{\citenamefont {McMahon}\ \emph {et~al.}(2016)\citenamefont
  {McMahon}, \citenamefont {Marandi}, \citenamefont {Haribara}, \citenamefont
  {Hamerly}, \citenamefont {Langrock}, \citenamefont {Tamate}, \citenamefont
  {Inagaki}, \citenamefont {Takesue}, \citenamefont {Utsunomiya}, \citenamefont
  {Aihara}, \citenamefont {Byer}, \citenamefont {Fejer}, \citenamefont
  {Mabuchi},\ and\ \citenamefont {Yamamoto}}]{YamSci}%
  \BibitemOpen
  \bibfield  {author} {\bibinfo {author} {\bibfnamefont {P.~L.}\ \bibnamefont
  {McMahon}}, \bibinfo {author} {\bibfnamefont {A.}~\bibnamefont {Marandi}},
  \bibinfo {author} {\bibfnamefont {Y.}~\bibnamefont {Haribara}}, \bibinfo
  {author} {\bibfnamefont {R.}~\bibnamefont {Hamerly}}, \bibinfo {author}
  {\bibfnamefont {C.}~\bibnamefont {Langrock}}, \bibinfo {author}
  {\bibfnamefont {S.}~\bibnamefont {Tamate}}, \bibinfo {author} {\bibfnamefont
  {T.}~\bibnamefont {Inagaki}}, \bibinfo {author} {\bibfnamefont
  {H.}~\bibnamefont {Takesue}}, \bibinfo {author} {\bibfnamefont
  {S.}~\bibnamefont {Utsunomiya}}, \bibinfo {author} {\bibfnamefont
  {K.}~\bibnamefont {Aihara}}, \bibinfo {author} {\bibfnamefont {R.~L.}\
  \bibnamefont {Byer}}, \bibinfo {author} {\bibfnamefont {M.~M.}\ \bibnamefont
  {Fejer}}, \bibinfo {author} {\bibfnamefont {H.}~\bibnamefont {Mabuchi}}, \
  and\ \bibinfo {author} {\bibfnamefont {Y.}~\bibnamefont {Yamamoto}},\ }\href
  {\doibase 10.1126/science.aah5178} {\bibfield  {journal} {\bibinfo  {journal}
  {Science}\ }\textbf {\bibinfo {volume} {354}},\ \bibinfo {pages} {614}
  (\bibinfo {year} {2016})}\BibitemShut {NoStop}%
\bibitem [{\citenamefont {Traversa}\ \emph
  {et~al.}(2015{\natexlab{a}})\citenamefont {Traversa}, \citenamefont
  {Ramella}, \citenamefont {Bonani},\ and\ \citenamefont
  {Ventra}}]{MEM_NP_complete_problems}%
  \BibitemOpen
  \bibfield  {author} {\bibinfo {author} {\bibfnamefont {F.~L.}\ \bibnamefont
  {Traversa}}, \bibinfo {author} {\bibfnamefont {C.}~\bibnamefont {Ramella}},
  \bibinfo {author} {\bibfnamefont {F.}~\bibnamefont {Bonani}}, \ and\ \bibinfo
  {author} {\bibfnamefont {M.~D.}\ \bibnamefont {Ventra}},\ }\href@noop {}
  {\bibfield  {journal} {\bibinfo  {journal} {Science Advances}\ }\textbf
  {\bibinfo {volume} {1}},\ \bibinfo {pages} {e1500031} (\bibinfo {year}
  {2015}{\natexlab{a}})}\BibitemShut {NoStop}%
\bibitem [{\citenamefont {Kowalsky}\ \emph {et~al.}(2022)\citenamefont
  {Kowalsky}, \citenamefont {Albash}, \citenamefont {Hen},\ and\ \citenamefont
  {Lidar}}]{3-xor-sat}%
  \BibitemOpen
  \bibfield  {author} {\bibinfo {author} {\bibfnamefont {M.}~\bibnamefont
  {Kowalsky}}, \bibinfo {author} {\bibfnamefont {T.}~\bibnamefont {Albash}},
  \bibinfo {author} {\bibfnamefont {I.}~\bibnamefont {Hen}}, \ and\ \bibinfo
  {author} {\bibfnamefont {D.~A.}\ \bibnamefont {Lidar}},\ }\href {\doibase
  10.1088/2058-9565/ac4d1b} {\bibfield  {journal} {\bibinfo  {journal} {Quantum
  Science and Technology}\ }\textbf {\bibinfo {volume} {7}},\ \bibinfo {pages}
  {025008} (\bibinfo {year} {2022})}\BibitemShut {NoStop}%
\bibitem [{\citenamefont {Mohseni}\ \emph {et~al.}(2022)\citenamefont
  {Mohseni}, \citenamefont {McMahon},\ and\ \citenamefont
  {Byrnes}}]{McMahon_Ising_machines}%
  \BibitemOpen
  \bibfield  {author} {\bibinfo {author} {\bibfnamefont {N.}~\bibnamefont
  {Mohseni}}, \bibinfo {author} {\bibfnamefont {P.~L.}\ \bibnamefont
  {McMahon}}, \ and\ \bibinfo {author} {\bibfnamefont {T.}~\bibnamefont
  {Byrnes}},\ }\href@noop {} {\bibfield  {journal} {\bibinfo  {journal} {Nature
  Reviews Physics}\ ,\ \bibinfo {pages} {1}} (\bibinfo {year}
  {2022})}\BibitemShut {NoStop}%
\bibitem [{\citenamefont {Lucas}(2014)}]{Lucas}%
  \BibitemOpen
  \bibfield  {author} {\bibinfo {author} {\bibfnamefont {A.}~\bibnamefont
  {Lucas}},\ }\href {\doibase 10.3389/fphy.2014.00005} {\bibfield  {journal}
  {\bibinfo  {journal} {Frontiers in Physics}\ }\textbf {\bibinfo {volume}
  {2}},\ \bibinfo {pages} {5} (\bibinfo {year} {2014})}\BibitemShut {NoStop}%
\bibitem [{\citenamefont {Barthel}\ \emph {et~al.}(2002)\citenamefont
  {Barthel}, \citenamefont {Hartmann}, \citenamefont {Leone}, \citenamefont
  {Ricci-Tersenghi}, \citenamefont {Weigt},\ and\ \citenamefont
  {Zecchina}}]{Hiding_Solutions}%
  \BibitemOpen
  \bibfield  {author} {\bibinfo {author} {\bibfnamefont {W.}~\bibnamefont
  {Barthel}}, \bibinfo {author} {\bibfnamefont {A.~K.}\ \bibnamefont
  {Hartmann}}, \bibinfo {author} {\bibfnamefont {M.}~\bibnamefont {Leone}},
  \bibinfo {author} {\bibfnamefont {F.}~\bibnamefont {Ricci-Tersenghi}},
  \bibinfo {author} {\bibfnamefont {M.}~\bibnamefont {Weigt}}, \ and\ \bibinfo
  {author} {\bibfnamefont {R.}~\bibnamefont {Zecchina}},\ }\href {\doibase
  10.1103/PhysRevLett.88.188701} {\bibfield  {journal} {\bibinfo  {journal}
  {Phys. Rev. Lett.}\ }\textbf {\bibinfo {volume} {88}},\ \bibinfo {pages}
  {188701} (\bibinfo {year} {2002})}\BibitemShut {NoStop}%
\bibitem [{\citenamefont {Ricci-Tersenghi}(2010)}]{Being_Glassy}%
  \BibitemOpen
  \bibfield  {author} {\bibinfo {author} {\bibfnamefont {F.}~\bibnamefont
  {Ricci-Tersenghi}},\ }\href {\doibase 10.1126/science.1189804} {\bibfield
  {journal} {\bibinfo  {journal} {Science}\ }\textbf {\bibinfo {volume}
  {330}},\ \bibinfo {pages} {1639} (\bibinfo {year} {2010})}\BibitemShut
  {NoStop}%
\bibitem [{\citenamefont {Hen}(2019)}]{PhysRevApplied.12.011003}%
  \BibitemOpen
  \bibfield  {author} {\bibinfo {author} {\bibfnamefont {I.}~\bibnamefont
  {Hen}},\ }\href {\doibase 10.1103/PhysRevApplied.12.011003} {\bibfield
  {journal} {\bibinfo  {journal} {Phys. Rev. Applied}\ }\textbf {\bibinfo
  {volume} {12}},\ \bibinfo {pages} {011003} (\bibinfo {year}
  {2019})}\BibitemShut {NoStop}%
\bibitem [{\citenamefont {Tsukamoto}\ \emph
  {et~al.}(2017{\natexlab{b}})\citenamefont {Tsukamoto}, \citenamefont
  {Takatsu}, \citenamefont {Matsubara},\ and\ \citenamefont {Tamura}}]{fuj1}%
  \BibitemOpen
  \bibfield  {author} {\bibinfo {author} {\bibfnamefont {S.}~\bibnamefont
  {Tsukamoto}}, \bibinfo {author} {\bibfnamefont {M.}~\bibnamefont {Takatsu}},
  \bibinfo {author} {\bibfnamefont {S.}~\bibnamefont {Matsubara}}, \ and\
  \bibinfo {author} {\bibfnamefont {H.}~\bibnamefont {Tamura}},\ }\href@noop {}
  {\bibfield  {journal} {\bibinfo  {journal} {Fujitsu Sci. Tech. J}\ }\textbf
  {\bibinfo {volume} {53}},\ \bibinfo {pages} {8} (\bibinfo {year}
  {2017}{\natexlab{b}})}\BibitemShut {NoStop}%
\bibitem [{\citenamefont {Johnson}\ \emph {et~al.}(2010)\citenamefont
  {Johnson}, \citenamefont {Bunyk}, \citenamefont {Maibaum}, \citenamefont
  {Tolkacheva}, \citenamefont {Berkley}, \citenamefont {Chapple}, \citenamefont
  {Harris}, \citenamefont {Johansson}, \citenamefont {Lanting}, \citenamefont
  {Perminov}, \citenamefont {Ladizinsky}, \citenamefont {Oh},\ and\
  \citenamefont {Rose}}]{dwave1}%
  \BibitemOpen
  \bibfield  {author} {\bibinfo {author} {\bibfnamefont {M.~W.}\ \bibnamefont
  {Johnson}}, \bibinfo {author} {\bibfnamefont {P.}~\bibnamefont {Bunyk}},
  \bibinfo {author} {\bibfnamefont {F.}~\bibnamefont {Maibaum}}, \bibinfo
  {author} {\bibfnamefont {E.}~\bibnamefont {Tolkacheva}}, \bibinfo {author}
  {\bibfnamefont {A.~J.}\ \bibnamefont {Berkley}}, \bibinfo {author}
  {\bibfnamefont {E.~M.}\ \bibnamefont {Chapple}}, \bibinfo {author}
  {\bibfnamefont {R.}~\bibnamefont {Harris}}, \bibinfo {author} {\bibfnamefont
  {J.}~\bibnamefont {Johansson}}, \bibinfo {author} {\bibfnamefont
  {T.}~\bibnamefont {Lanting}}, \bibinfo {author} {\bibfnamefont
  {I.}~\bibnamefont {Perminov}}, \bibinfo {author} {\bibfnamefont
  {E.}~\bibnamefont {Ladizinsky}}, \bibinfo {author} {\bibfnamefont
  {T.}~\bibnamefont {Oh}}, \ and\ \bibinfo {author} {\bibfnamefont
  {G.}~\bibnamefont {Rose}},\ }\href {\doibase 10.1088/0953-2048/23/6/065004}
  {\bibfield  {journal} {\bibinfo  {journal} {Superconductor Science and
  Technology}\ }\textbf {\bibinfo {volume} {23}},\ \bibinfo {pages} {065004}
  (\bibinfo {year} {2010})}\BibitemShut {NoStop}%
\bibitem [{\citenamefont {Bernaschi}\ \emph {et~al.}(2021)\citenamefont
  {Bernaschi}, \citenamefont {Bisson}, \citenamefont {Fatica}, \citenamefont
  {Marinari}, \citenamefont {Martin-Mayor}, \citenamefont {Parisi},\ and\
  \citenamefont {Ricci-Tersenghi}}]{SATonGPU}%
  \BibitemOpen
  \bibfield  {author} {\bibinfo {author} {\bibfnamefont {M.}~\bibnamefont
  {Bernaschi}}, \bibinfo {author} {\bibfnamefont {M.}~\bibnamefont {Bisson}},
  \bibinfo {author} {\bibfnamefont {M.}~\bibnamefont {Fatica}}, \bibinfo
  {author} {\bibfnamefont {E.}~\bibnamefont {Marinari}}, \bibinfo {author}
  {\bibfnamefont {V.}~\bibnamefont {Martin-Mayor}}, \bibinfo {author}
  {\bibfnamefont {G.}~\bibnamefont {Parisi}}, \ and\ \bibinfo {author}
  {\bibfnamefont {F.}~\bibnamefont {Ricci-Tersenghi}},\ }\href {\doibase
  10.1209/0295-5075/133/60005} {\bibfield  {journal} {\bibinfo  {journal}
  {Europhysics Letters}\ }\textbf {\bibinfo {volume} {133}},\ \bibinfo {pages}
  {60005} (\bibinfo {year} {2021})}\BibitemShut {NoStop}%
\bibitem [{\citenamefont {Traversa}\ \emph
  {et~al.}(2015{\natexlab{b}})\citenamefont {Traversa}, \citenamefont
  {Ramella}, \citenamefont {Bonani},\ and\ \citenamefont {Ventra}}]{MEM}%
  \BibitemOpen
  \bibfield  {author} {\bibinfo {author} {\bibfnamefont {F.~L.}\ \bibnamefont
  {Traversa}}, \bibinfo {author} {\bibfnamefont {C.}~\bibnamefont {Ramella}},
  \bibinfo {author} {\bibfnamefont {F.}~\bibnamefont {Bonani}}, \ and\ \bibinfo
  {author} {\bibfnamefont {M.~D.}\ \bibnamefont {Ventra}},\ }\href {\doibase
  10.1126/sciadv.1500031} {\bibfield  {journal} {\bibinfo  {journal} {Science
  Advances}\ }\textbf {\bibinfo {volume} {1}},\ \bibinfo {pages} {e1500031}
  (\bibinfo {year} {2015}{\natexlab{b}})}\BibitemShut {NoStop}%
\bibitem [{\citenamefont {Goto}\ \emph
  {et~al.}(2019{\natexlab{b}})\citenamefont {Goto}, \citenamefont {Tatsumura},\
  and\ \citenamefont {Dixon}}]{sbm1}%
  \BibitemOpen
  \bibfield  {author} {\bibinfo {author} {\bibfnamefont {H.}~\bibnamefont
  {Goto}}, \bibinfo {author} {\bibfnamefont {K.}~\bibnamefont {Tatsumura}}, \
  and\ \bibinfo {author} {\bibfnamefont {A.~R.}\ \bibnamefont {Dixon}},\ }\href
  {\doibase 10.1126/sciadv.aav2372} {\bibfield  {journal} {\bibinfo  {journal}
  {Science Advances}\ }\textbf {\bibinfo {volume} {5}},\ \bibinfo {pages}
  {eaav2372} (\bibinfo {year} {2019}{\natexlab{b}})}\BibitemShut {NoStop}%
\bibitem [{\citenamefont {Swendsen}\ and\ \citenamefont {Wang}(1986)}]{PT1}%
  \BibitemOpen
  \bibfield  {author} {\bibinfo {author} {\bibfnamefont {R.~H.}\ \bibnamefont
  {Swendsen}}\ and\ \bibinfo {author} {\bibfnamefont {J.-S.}\ \bibnamefont
  {Wang}},\ }\href {\doibase 10.1103/PhysRevLett.57.2607} {\bibfield  {journal}
  {\bibinfo  {journal} {Phys. Rev. Lett.}\ }\textbf {\bibinfo {volume} {57}},\
  \bibinfo {pages} {2607} (\bibinfo {year} {1986})}\BibitemShut {NoStop}%
\bibitem [{\citenamefont {Hukushima}\ and\ \citenamefont {Nemoto}(1996)}]{PT2}%
  \BibitemOpen
  \bibfield  {author} {\bibinfo {author} {\bibfnamefont {K.}~\bibnamefont
  {Hukushima}}\ and\ \bibinfo {author} {\bibfnamefont {K.}~\bibnamefont
  {Nemoto}},\ }\href {\doibase 10.1143/JPSJ.65.1604} {\bibfield  {journal}
  {\bibinfo  {journal} {Journal of the Physical Society of Japan}\ }\textbf
  {\bibinfo {volume} {65}},\ \bibinfo {pages} {1604} (\bibinfo {year}
  {1996})}\BibitemShut {NoStop}%
\bibitem [{\citenamefont {Rønnow}\ \emph {et~al.}(2014)\citenamefont
  {Rønnow}, \citenamefont {Wang}, \citenamefont {Job}, \citenamefont {Boixo},
  \citenamefont {Isakov}, \citenamefont {Wecker}, \citenamefont {Martinis},
  \citenamefont {Lidar},\ and\ \citenamefont {Troyer}}]{TTS}%
  \BibitemOpen
  \bibfield  {author} {\bibinfo {author} {\bibfnamefont {T.~F.}\ \bibnamefont
  {Rønnow}}, \bibinfo {author} {\bibfnamefont {Z.}~\bibnamefont {Wang}},
  \bibinfo {author} {\bibfnamefont {J.}~\bibnamefont {Job}}, \bibinfo {author}
  {\bibfnamefont {S.}~\bibnamefont {Boixo}}, \bibinfo {author} {\bibfnamefont
  {S.~V.}\ \bibnamefont {Isakov}}, \bibinfo {author} {\bibfnamefont
  {D.}~\bibnamefont {Wecker}}, \bibinfo {author} {\bibfnamefont {J.~M.}\
  \bibnamefont {Martinis}}, \bibinfo {author} {\bibfnamefont {D.~A.}\
  \bibnamefont {Lidar}}, \ and\ \bibinfo {author} {\bibfnamefont
  {M.}~\bibnamefont {Troyer}},\ }\href {\doibase 10.1126/science.1252319}
  {\bibfield  {journal} {\bibinfo  {journal} {Science}\ }\textbf {\bibinfo
  {volume} {345}},\ \bibinfo {pages} {420} (\bibinfo {year}
  {2014})}\BibitemShut {NoStop}%
\end{thebibliography}%

\end{document}